\documentclass{eptcs}
\providecommand{\event}{ADG 2021} 
\usepackage{graphicx}
\usepackage{amssymb}
\usepackage{epstopdf}
\usepackage{subfig}
\usepackage[utf8]{inputenc}
\usepackage{pgf,tikz}
\usetikzlibrary{arrows}
\usepackage{color,xcolor}
\usepackage{amsmath}
\def\orcidID#1{\unskip$^{\mbox{\href{https://orcid.org/#1}{\scriptsize{[#1]}} }}$}

\graphicspath{{./figs/}}

\newcommand{\OO}{\mathcal{O}}

\newcommand{\bO}{\mathbf{O}} 

\newcommand{\afold}{\looparrowright}
\newcommand{\sfold}{\looparrowright} 

\newcommand{\eos}{\textsc{Eos}}

\newcommand{\squeezelist}{
 \begin{list}{$\bullet$}
  { \setlength{\itemsep}{0pt}
     \setlength{\leftmargin}{1.2em}
      } }
\newcommand{\squeezelisttwo}{
 \begin{list}{$\bullet$}
  {
     \setlength{\leftmargin}{1.2em}
      }}

\newcommand{\defeq}{{\triangleq}}
\newcommand{\spp}{\succ}
\newcommand{\adj}{\backsim}

\newcommand{\comment}[1]{}

\begin{document}
\title{A New Modeling of Classical Folds in Computational 
Origami\thanks{Supported by JSPS KAKENHI Grant No.16K00008}}

\author{Tetsuo Ida\orcidID{0000-0002-5683-216X} 
\institute{University of Tsukuba \\ 
Tsukuba 305-8573, Japan} \\
\url{https://www.i-eos.org/ida} 
\email{ida@cs.tsukuba.ac.jp}
\and
Hidekazu Takahashi\orcidID{0000-0002-2866-8876} 
\institute{Hikone Higashi High School \\
Hikone 522-0061, Japan}
\email{t.hidekazu@gmail.com}
}

\def\titlerunning{A New Modeling of Classical Folds in Computational Origami}
\def\authorrunning{Ida \& Takahashi}

\maketitle

\begin{abstract}
This paper shows a cut along a crease on an origami sheet makes simple modeling of popular traditional basic folds such as a squash fold in computational origami. The cut operation can be applied to other classical folds and significantly simplify their modeling and subsequent implementation in the context of computational origami.
\end{abstract}


\section{Introduction}
Origami attracts the minds of people all over the world. Some are interested in its geometric aspects, and others in artistic or recreational elements, so-called \emph{traditional origami}. Although both origami categories rely on a single notion of paper folding, their methodologies differ significantly due to the pursuits' objectives and goals. In this paper, we focus on traditional origami.

We can find numerous origami artworks in web pages, books and technical papers; some are in the realm of artistic creation, and others are pieces of origami works for recreation and educational purposes.  The origami creators' interest lies more in artistic creativity than general folding rules in the first category. In contrast, in the second category, the underlying motivation is to find a finite number of basic fold rules.  Although in 2D (two dimensional) origami geometry, we can obtain such a collection of rules, in traditional origami, we are more liberal to collect fold rules since the set may be open.  Nevertheless, we give some basic fold rules that we use in traditional origami.  We start from the description of the web pages~\cite{corsoyard:2021}.

We understand the 2D origami geometry to the extent that we do the 2D Euclidean geometry in terms of the constructible points in the two geometries. Namely, the set of the points constructible by Huzita-Justin's fold rules~\cite{Huzita:1989a,Justin:1986} is the strict superset of intersecting points of the circles and the lines made by a straightedge and a compass - the construction tool of Euclidean geometry. Alperin gave a more concise description of this fact in the language of  fields~\cite{Alperin:2000}.

However, we do not have a clear view of formalization of traditional origami. The formalization of traditional origami, in its entirety, is not our intended goal. Instead, we prefer to model some of the crucial and sophisticated operations using the results of the accumulated formalization efforts~(e.g., \cite{Lang:2003,folds-net:2005,corsoyard:2021}).


 In our earlier work, we presented an abstract origami 
 system~\cite{ida:2010a}. It involves the notions of two relations on the set of origami faces. By them, we can model the superpositions of pieces of origami faces. The superpositions of the faces make origami look like a three-dimensional (3D) object. To superpose faces intriguingly makes the origami artistic or a thing that many people fancy to construct. Classical folds have been discovered and elaborated over time to the present day to cater to these desires.
For example, on the web pages of corsoyard.com, they show more than ten classical folding techniques~\cite{corsoyard:2021}. As typified by the squash fold and inside-reverse fold, the classical fold intricately combines several simple folds. One squash fold, for example, requires the simultaneous three simple folds, i.e., mountain or valley fold, subjected to non-trivial constraints.   The combination of those three folds will not afford us simple modeling of the fold.

This paper shows a cut along a crease makes simple modeling of a squash fold in computational origami. The cut operation can be applied to other classical folds and significantly simplify their modeling and subsequent implementation in the context of computational origami.
\section{Preliminary}
Let $\Pi$ be a finite set of (origami) faces, $\adj$ be a binary relation on $\Pi$, called \emph{adjacency relation} and $\spp$ be a binary relation on $\Pi$, called \emph{superposition relation}. An abstract origami is a structure $(\Pi, \adj, \spp)$.
We abbreviate abstract origami to \emph{AO}.
We denote the set of AOs by $\bO$.
An abstract origami system is an abstract
rewriting system $(\bO, \sfold)$~\cite{klop:2003} , where 
$\sfold$ is a rewrite relation on $\bO$, called \emph{abstract fold}.

For $\OO, \OO' (\in \bO)$, we write $\OO
\sfold \OO'$ when $\OO$ is abstractly folded to $\OO'$. 
We start an origami construction with an initial AO
and perform an abstract fold repeatedly until we obtain the desired AO. Usually, we start an origami construction with a square sheet of paper. This initial sheet of paper is abstracted as a structure having a single distinguished face denoted by the numeral $1$. Then, the initial AO $\OO_{1}$ is represented by $(\{1\}, \emptyset, \emptyset)$.  We could use any symbol to denote the initial face. Still, we find it convenient to use the numeral '$1$' as we can interpret it as a numeric value to generate the denotation of the new faces during the construction.  We will see in the concrete examples that when we fold face $n$, the face is divided into two faces $2n$ and $2n+1$.  We use this convention in this paper and the realization of the data structure of the computational origami system \eos~\cite{ida:2004}. 
Suppose that we are at the beginning of step $i$ of the construction, having AO $\OO_{i-1}=(\Pi_{i-1}, \adj_{i-1}, \spp_{i-1})$. We perform an abstract fold and
obtain a next AO $\OO_{i}=(\Pi_{i}, \adj_{i}, \spp_{i})$. Thus, we have the
following $\sfold$-sequence.
\[ \OO_1 \sfold \OO_2\afold \cdots \sfold \OO_n\]
An abstract origami construction is a finite $\sfold$-sequence of AOs.

In concrete terms, the operation $\sfold$ can be a Huzita-Justin's fold rule, a mountain fold, a valley fold, etc., each requiring different kinds of arguments. For further details, refer to~\cite{ida:2020}, Chapter 7.
\begin{figure}[ht]
\begin{center}
\subfloat[origami house in 2D]
{\includegraphics[scale=1.0]{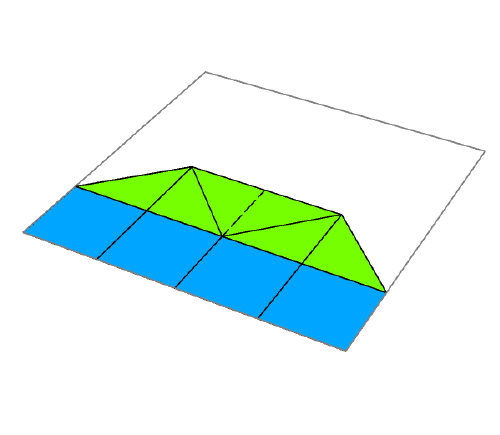}}
\hfil
\subfloat[origami house in 3D]
{\includegraphics[scale=1.5]{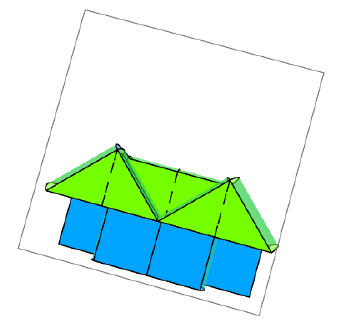}}
\caption{Origami house}
\label{fig:origami-house}
\end{center}
\end{figure}

\section{A motivating example}
We consider a squash fold since it is used frequently in traditional origami. See, for example, a popular book on origami~\cite{ninkino-origami:2011}.  

\begin{figure}[h]
\subfloat[$\OO_1$]{\includegraphics[scale=0.87]{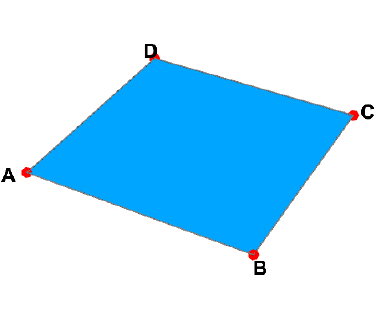}\label{h-step1}}
\quad
\subfloat[$\OO_2$]{\includegraphics[scale=0.87]{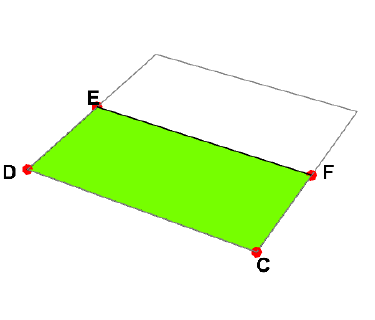}}
\quad
\subfloat[$\OO_4$]{\includegraphics[scale=0.87]{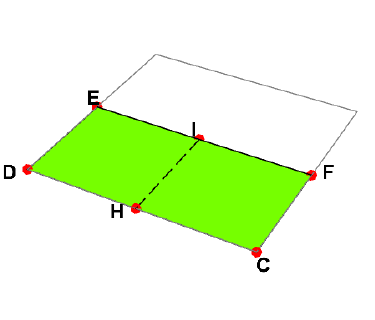}}
\quad
\subfloat[$\OO_6$]{\includegraphics[scale=0.87]{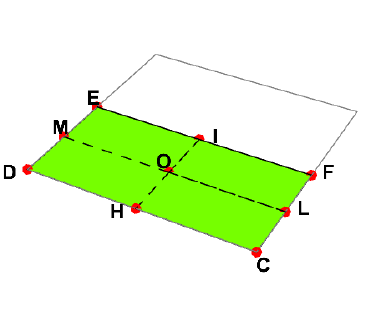}}
\quad
\subfloat[$\OO_7$]{\includegraphics[scale=0.87]{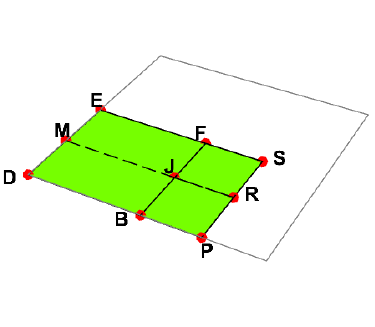}}
\caption{House onstruction sequence: selected AOs}
\label{fig:house-construction}
\end{figure}

\begin{figure}[h]
\begin{center}
\subfloat[$\OO_9$: before the squash fold]
{\includegraphics[width=50mm,height=40mm]{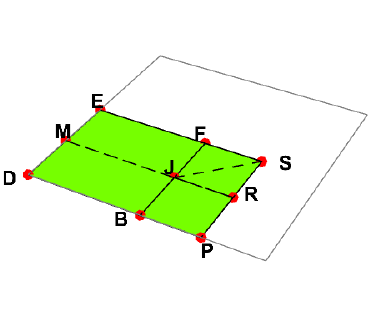}}
\quad
\subfloat[$\OO_{14}$: after the squash fold]
{\includegraphics[width=50mm,height=40mm]{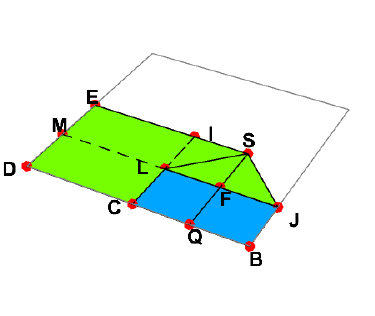}}
\end{center}
\caption{Squash fold in origami house construction}
\label{fig:squash-fold-origami-house}
\end{figure}

Figure~\ref{fig:origami-house} shows an origami house that most children can compose when they learn the recipe of the origami house. The only non-trivial steps are the squash folds on the right- and left-hand roofs of the house.
Figures~\ref{fig:house-construction} and~\ref{fig:squash-fold-origami-house} show the selected AOs in the construction sequence. When we show AOs in the figures, we show their geometrical interpretation rather than express them symbolically.


We briefly explain how we obtain each AO in the sequence.
\begin{itemize}
 \item[]
Step 1: We start with the initial origami $\OO_1$, which we represent  geometrically  in \ref{fig:house-construction}(a).
 \item[]
Step 2: Fold $\OO_1$ to bring D to A in Fig.~\ref{fig:house-construction}(b).
 \item[]
Steps 3 and 4: Fold $\OO_2$ to bring D to C, and then unfold  $\OO_3$ in Fig.~\ref{fig:house-construction}(c).
 \item[]
Steps 5 and 6: Fold $\OO_4$ to bring F to C,  and then unfold  $\OO_5$ in Fig.~\ref{fig:house-construction}(d).
 \item[]
Step 7: Fold $\OO_6$ to bring C to H in Fig.~\ref{fig:house-construction}(e).
 \item[]
Steps 8 and 9: Fold $\OO_7$ to bring F to R,  and then unfold  $\OO_8$ in Fig.~\ref{fig:squash-fold-origami-house}(a).
\end{itemize}

From the construction sequence, we see a slit below faces JSF and SJR, i.e., square JRSF in Fig.~\ref{fig:squash-fold-origami-house}(a). We pull point~J slightly upward and make a small space below JRSF. Then, we move segments SF and SJ with S fixed, such that SF overlays SR. By this movement, we squash face JSF onto the area to the right of segment SR. After the squash, the rotated segment SF and non-moved segment SR, as a whole, look like a ridge and a valley that departs at S. This fold invokes three simultaneous face moves by three fold lines. It is a sophisticated motion, and none of Huzita-Justin rules can cope with it. In applying Huzita-Justin rules, we can employ only one fold line in a single fold.
If we allow a cut along crease SJ, we simplify the moves of the involved faces. After we complete this squash operation, we glue the edges separated by the cut operation. The algorithm of \emph{squash fold with a cut} in the next section makes clear the involved face moves.

To complete the construction of the origami house, we perform the same sequence of fold operation on the left part of the rectangle DBFE, although in a mirror fashion across line IC.

\section{Squash fold with a cut}
\begin{figure}[h]
\subfloat[$\OO_1$]{\includegraphics[scale=0.65]{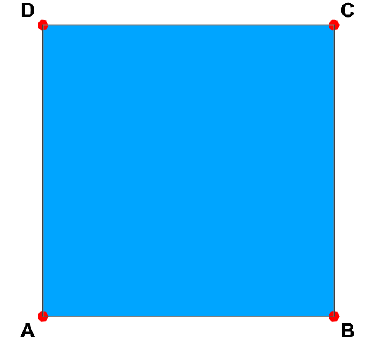}}
\subfloat[$\OO_2$]{\includegraphics[scale=0.65]{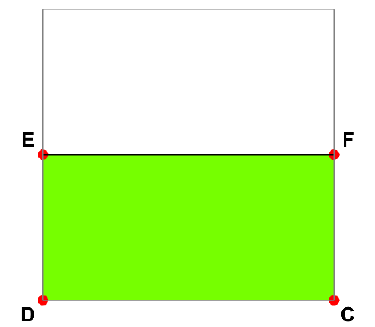}}
\subfloat[$\OO_3$]{\includegraphics[scale=0.65]{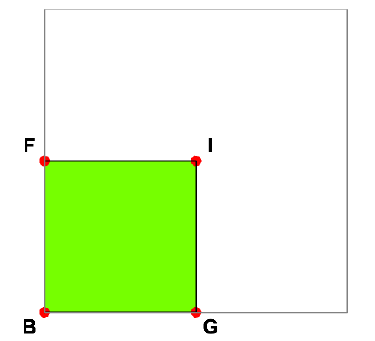}}
\subfloat[$\OO_4$]{\includegraphics[scale=0.65]{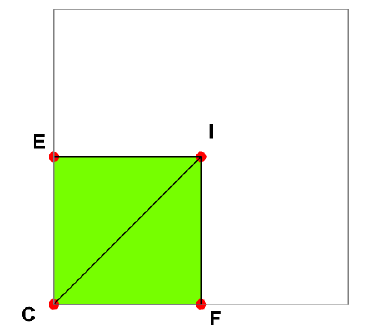}}
\subfloat[$\OO_5$]{\includegraphics[scale=0.65]{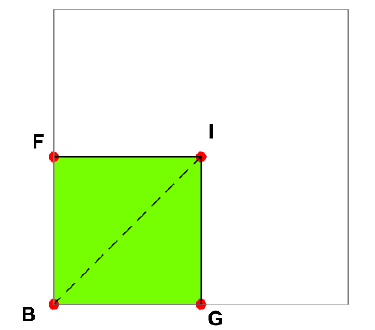}}

\caption{Construction sequence $\OO_1 \sfold^{+}\OO_5$}
\label{fig:squash-fold-origami-before}
\end{figure}

\begin{figure}[h]
\begin{center}
\includegraphics[scale=1.5]{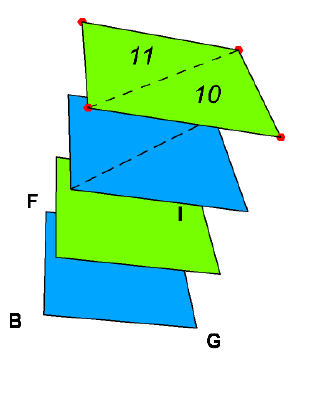}
\end{center}
\caption{Visualization of $\OO_5 =(\Pi_5, \adj_5, \spp_5)$}
\label{fig:squash-step51}
\end{figure}

\begin{figure}[h]
\center
\subfloat[Adjacency graph $(\Pi_5, \adj_5)$]
{\includegraphics[scale=0.7]{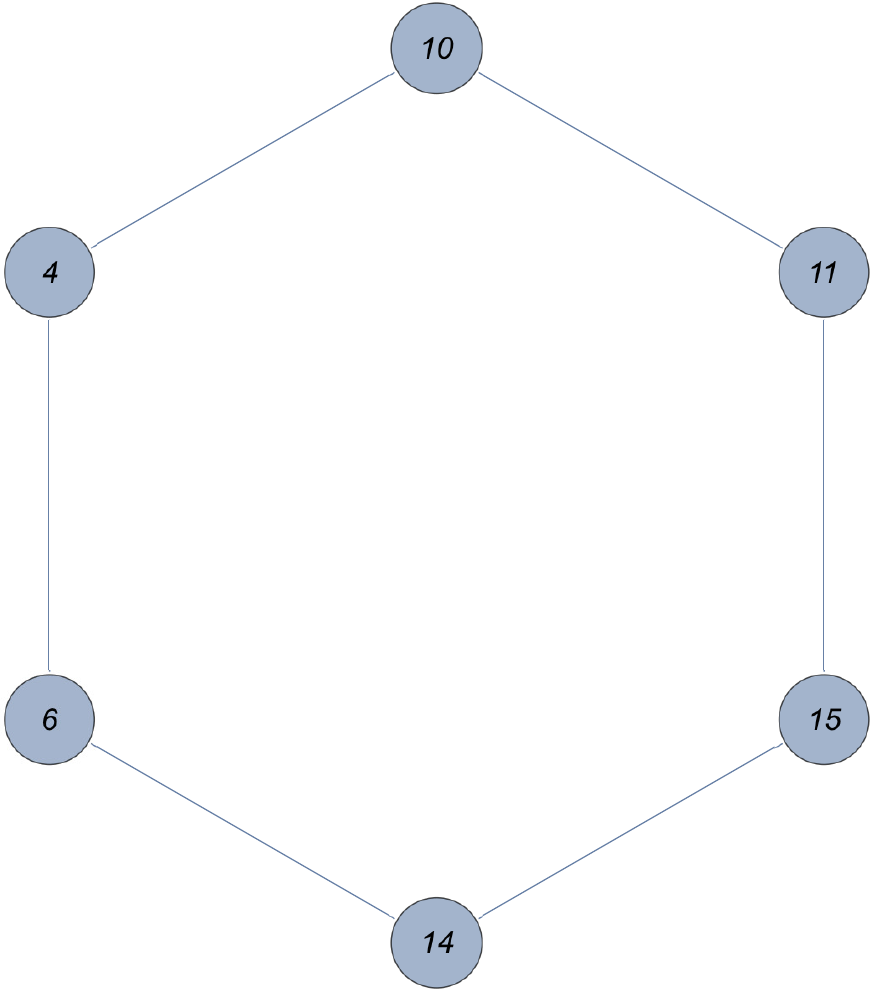}}
\qquad\qquad
\subfloat[Superposition graph $(\Pi_5, \spp_5)$]
{\includegraphics[scale=0.8]{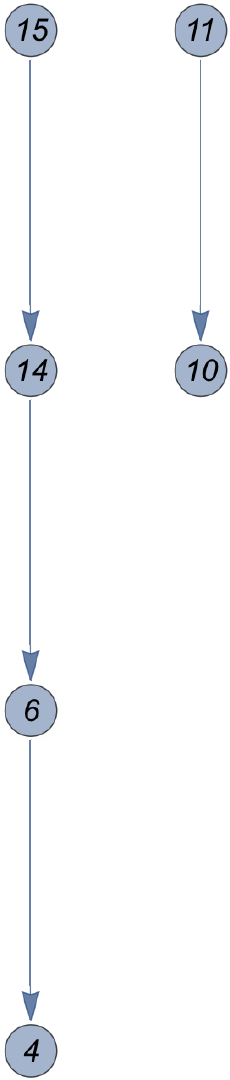}}
\caption{Graphs of $\OO_5 =(\Pi_5, \adj_5, \spp_5)$}
\label {fig:O5-graph1}
\end{figure}

In its simplest form, we perform a squash fold on a four-layered square  $\OO_5$ shown in Fig. ~\ref{fig:squash-fold-origami-before}(e). 
We can visualize  $\OO_5$ more clearly by expanding the vertical gap between the faces, as in Fig.~\ref{fig:squash-step51}.  
We have $\Pi_5 =\{4, 6, 10, 11, 14 ,15\}$. We could represent the relations $\adj_5$ and $\spp_5$ by the usual notation of a set of pairs of the denotations of the faces in $\Pi_5$, but to facilitate the analysis of the relations, we use the graphs of $(\Pi_5, \adj_5)$ and $(\Pi_5, \spp_5)$.  Figures~\ref{fig:O5-graph1}(a)
~and~\ref{fig:O5-graph1}(b) show the relations $\adj_5$ and $\spp_5$, respectively. We note that faces ${10} $ and ${11}$ are adjacent. We can also observe the adjacency of faces ${10}$ and ${11}$ in Figs.~\ref{fig:squash-fold-origami-before}(e) and \ref{fig:squash-step51}. The dotted line borders faces ${10}$ and ${11}$.  As a convention, a dotted line denotes a valley crease in our graphics visualization.

To reason about the geometric properties of 3D origami, we extensively use tools of graph theory and 3D visualizations.  We needed to develop tools of the latter by ourselves as part of \eos.
With this preparation, we now describe the algorithm of a squash fold with a cut.\\

\begin{figure}[h]
\center
\subfloat[$\OO_7$ at step 7]{\includegraphics[scale=0.9]{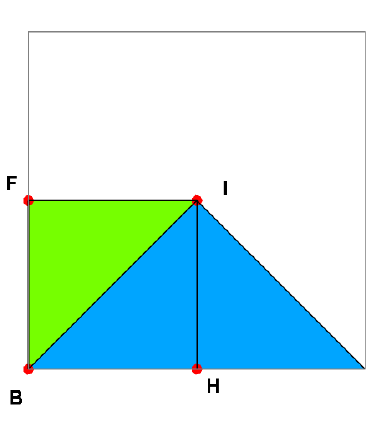}}
\quad
\subfloat[$\OO_8$ at step 8]{\includegraphics[scale=0.9]{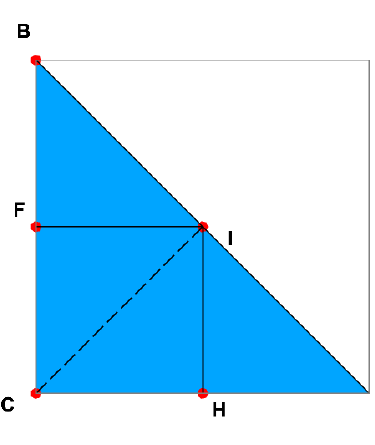}}
\quad
\subfloat[$\OO_9$ at step 9]{\includegraphics[scale=1.15]{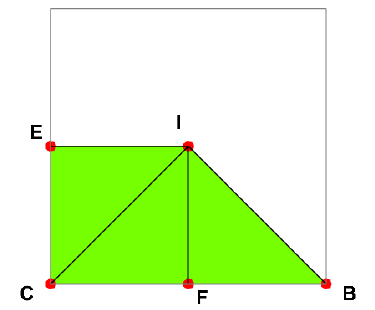}}
\caption{Folds in a squash fold}
\label{fig:squash-fold-process}
\end{figure}

\begin{figure}[ht]
\center
\subfloat[Adjacency graph $(\Pi_{10}, \adj_{10})$]{\includegraphics[scale=0.9]{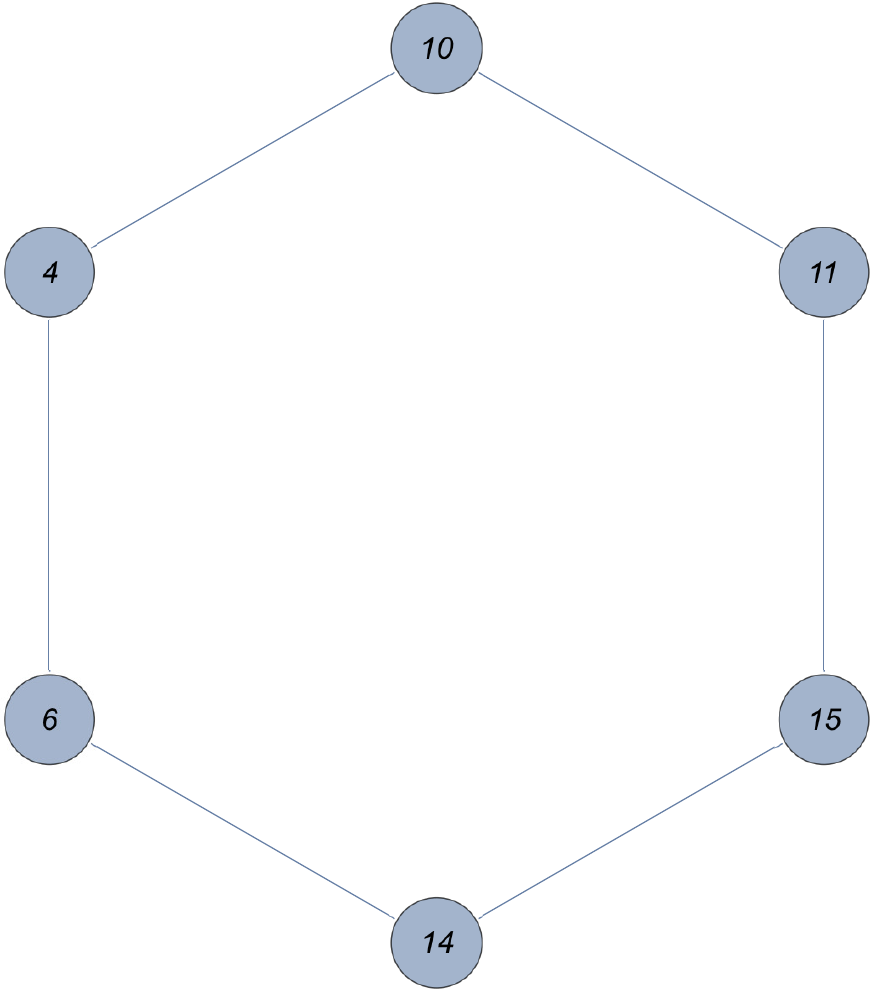}}
\qquad\qquad
\subfloat[Superposition graph $(\Pi_{10}, \spp_{10})$]{\includegraphics[scale=1.1]{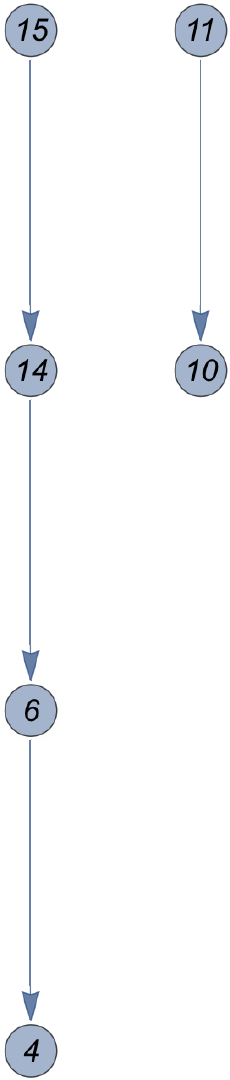}}

\caption{Graphs of origami after squash fold}
\label{fig:Graphs-of-O10}
\end{figure}

{\bf Algorithm} \textbf{Squash fold with a cut}
\begin{enumerate}
\item
Cut the edge between nodes ${10}$ and ${11}$ of the adjacency graph of 
Fig.~\ref{fig:O5-graph1}(a). 
The edge corresponds to the segment IB in Fig.~\ref{fig:squash-fold-origami-before}(e).
The cut enables the subsequent three folds. 
We save the encoding of the separated edges for later glue at Step 5.
\item
Fold face 10 along ray IG (cf. Fig.~\ref{fig:squash-fold-process}(a)).
\item 
Fold face 11 along ray FI (cf. Fig.~\ref{fig:squash-fold-process}(b)).
\item
Fold to bring point F to point G. Equivalently, fold along segment IC
 (cf. Fig.~\ref{fig:squash-fold-process}(c)).
\item
Glue the edge separated at Step 1, i.e.,  edge IB.
\end{enumerate}
We show the result of the execution of the algorithm in 3D, together with the associated
adjacency and superposition graphs in Fig.\ref{fig:Graphs-of-O10}.

\begin{figure}[h]
\begin{center}
\includegraphics[scale=1.5]{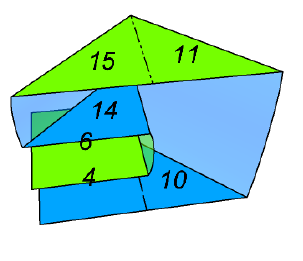}

The curved surfaces between the edges\\ of the two overlapping faces show that they are connected.
\caption{Visualization in 3D of $\OO_{10} =(\Pi_{10}, \adj_{10}, \spp_{10})$}
\end{center}
\label{fig:squash-step102}
\end{figure}

Once we design the algorithm of the squash fold, we can easily abstract the algorithm into a function called SquashFold.  Then, we treat the call of SquashFold with appropriate arguments as if the squash fold were a single operation.  We describe SquashFold in the syntax of  Wolfram Language of Mathematica~\cite{Wolfram:2021}.   Function SquashFold is incorporated into  \eos.  \verb|SquashFold0| below performs the algorithm.\footnote{We omit optional arguments in function definitions  for brevity.}  It is one of the rewrite rules that define function SquashFold.  We have other rules for different kinds of arguments of the squash fold.
\begin{verbatim}
SquashFold0[{below_,above_},{bottom:Ray[P_,Q_],
 ridge:Ray[R_,P_]}]:=
(CutEdge[{below,above}];
 ValleyFold[below,bottom];ValleyFold[above,ridge];HO[R,Q];
 GlueEdges[$cutEdge])
\end{verbatim}

The first  line and the second (to the left of \verb|:=|) line of the program is a left-hand side of a rewrite rule in which we specify input arguments; below and above are faces that will become below and above the faces of the squashed part of the origami.  In this case, the variables below and above are bound to 10 and 11, respectively,  when we execute the following function call.
\begin{center}
\begin{verbatim}
SquashuFold0[{10,11},{Ray["I","G"],Ray["F","I"]}]
\end{verbatim}
\end{center}
The other variables, i.e., variables P, Q, and R, are bound to \verb|"I"|, \verb|"G"|, and \verb|"F"|, respectively.  Note that we enclose point names with double quotation marks when we write \eos\ programs.

The third to fifth lines of the program are a constituent of the right-hand side of the rewrite rule, and their meanings are self-explanatory, except for the global variable \verb|$cutEdge|, which holds the encoding of the cut edges.  The fourth line is the program's nucleus.  It performs lines  2, 3, and 4 of Algorithm \textbf{Squash fold with a cut}. In the program, we evaluate the functions with specific names to its operations, i.e., functions corresponding to valley fold, valley fold and Huzita-Justin rule O2.

In summary, we have observed that at the level of abstract origami rewriting.
\[ 
\OO_{1} \sfold_{1} \OO_{2} \sfold_{2} \OO_{3} \sfold_{3} \OO_{4} \sfold_{4}\OO_{5} \sfold_{5} \OO_{6} \sfold_{6} \OO_{7} \sfold_{7} \OO_{8} \sfold_{8}\OO_{9} \sfold_{9} \OO_{10} 
\] 
We call
the composition of rewrites
 $\sfold_{9} \circ \sfold_{8} \circ \sfold_{7} \circ \sfold_{6}\circ \sfold_{5} $
a \emph{rewrite of a squash fold}.

We can also observe that the adjacency graphs of $\OO_{5}$ and $\OO_{10}$ are the same.

\section{Basic folds in traditional origami}
We remarked in the introduction that origamists are more interested in the final origami shapes than the rule set of folds in traditional origami practice. Nevertheless, it is convenient to have a small collection of basic folding rules. In this section, we present some candidates of such basic rules.  
\begin{itemize}
\item
Huzita-Justin rules (seven rules, O1 -- O7,) 
\item
Mountain fold and valley fold
\item
Inside-reverse fold
\item
Outside-reverse fold
\item
Rabbit-ear fold
\item
Pleat fold and crimp-pleat fold
\end{itemize}  
\subsection{Huzita-Justin rules}
Huzita and Justin proposed the rules in the context of 2D origami geometry. They are the alternatives to a compass and a straightedge of the 2D Euclidean geometry. The rules define a fold line, and we fold the origami along it. We use the fold line to perform either a mountain fold or a valley fold.  
We restrict the use of Huzita-Justin rules to geometric origami, primarily.  However, we have seen with Huzita-Justin rules we can construct a variety of origami works.

\subsection{Mountain fold and valley fold}
Mountain and valley folds are applicable, in \eos, to a set of the faces of the abstract origami. They can be multi-layered. \eos\ automatically computes the target faces on which it performs folds. Thus, Functions MountainFold and ValleyFold can take arguments of a list of origami faces and a fold line. We can perceive the difference between ``mountain'' and ``valley'' when we unfold the origami immediately after the fold. If we see the crease made by the fold line that looks like a valley, it is a valley fold; if the crease looks like a mountain ridge, it is a mountain fold.

These are the most basic fold operations. We note that fold lines have no association with directions. Therefore, we use a ray instead of a line when \eos\ cannot determine the direction of the face movement in folding by other parameters.  With the use of rays, we can observe the right-/left-hand side of the line. We adopt the right-handed system and move the faces to the right of the ray when we fold the origami along the ray(line).

In the program of \verb|SquashFold0|, we already used two valley folds and one Huzita-Justin rule O2. We may replace the rule O2 with a fold along a ray that forms the perpendicular bisector of Segment[R, Q]. However, the O2 rule is applicable in this case since the target faces together form a flat plane. Furthermore, it is easier to reason about the geometric properties with Huzita-Justin rules. 

\subsection{Inside reverse fold and outside reverse fold}

\begin{figure}[h]
\center
\subfloat[Before InsideReverseFold]{\includegraphics[scale=1.1]{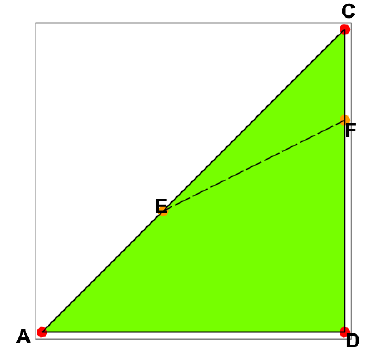}}
\qquad
\subfloat[After InsideReverseFold]{\includegraphics[scale=1.4]{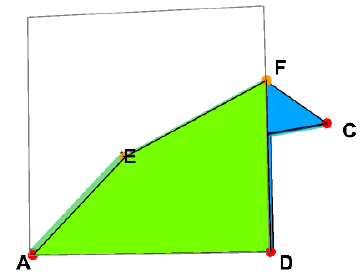}}
\caption{Inside reverse fold}
\label{fig:inside-reverse-fold}
\end{figure}

\begin{figure}[h]
\center
\subfloat[Before OutsideReverseFold]{\includegraphics[scale=1.1]{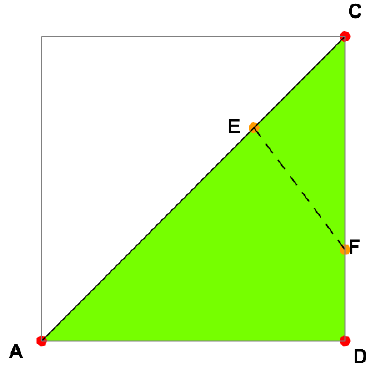}}
\qquad
\subfloat[After OutsideReverseFold]{\includegraphics[scale=1.4]{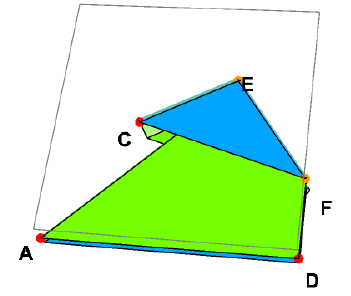}}
\caption{Outside reverse fold}
\label{fig:outside-reverse-fold}
\end{figure}

An inside reverse fold and an outside reverse fold are very similar operations. Suppose we have a pair of overlapping faces that share an edge. We fold the two faces by moving the part of the faces as in Fig.~\ref{fig:inside-reverse-fold} and Fig.~\ref{fig:outside-reverse-fold}. The figures clearly show what actions are involved. We need two (but the same) rays to perform mountain and valley folds simultaneously. The inside reverse and the outside reverse folds require two folds in a restricted way.  Hence, we see that we cannot apply Huzita-Justin rules. We can use the cut as in the squash fold. We show programs of the inside reverse fold and the outside reverse fold, too. They are easy to understand since we use the same set of functions similarly.

We obtain the origami shape in Fig.~\ref{fig:inside-reverse-fold} by executing:
\begin{verbatim}
InsideReverseFold[{5,7},Ray["F","E"]]
\end{verbatim}
where  function \verb|InsideReverseFold| is defined by
\begin{verbatim}
InsideReverseFold[{below_,above_},ray_] := 
 (CutEdge[{below,above}];
  MountainFold[above,ray,InsertFace->below]; 
  ValleyFold[below,ray,InsertFace->above];
  GlueEdges[$cutEdge])
  \end{verbatim}
 The numerals 5 and 7 of  the faces are obtained by the visualization tools, which are not discussed here.
  
Likewise, we obtain the origami in Fig.~\ref{fig:outside-reverse-fold} by executing:
\begin{verbatim}
InsideReverseFold[{5,7},Ray["F","E"]]
\end{verbatim}
where  function \verb|OutsideReverseFold| is defined by
\begin{verbatim}
OutsideReverseFold[{below_,above_},ray_] :=
 (CutEdge[{below,above}];
  ValleyFold[above,ray]; 
  MountainFold[below,ray]; 
  GlueEdges[$cutEdge])   
  \end{verbatim}

\subsection{Rabbit ear fold}
\begin{figure}[h]
\center
\subfloat[$\OO_{4}$]{\includegraphics[scale=1.1]{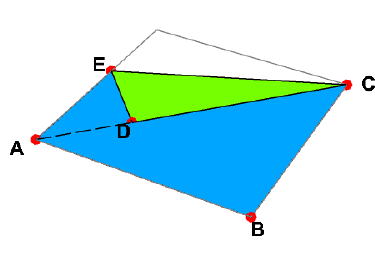}}
\quad
\subfloat[$\OO_{5}$]{\includegraphics[scale=1.1]{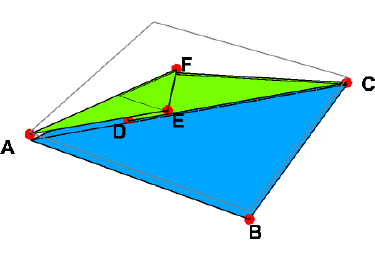}}
\quad
\subfloat[$\OO_{7}$]{\includegraphics[scale=1.1]{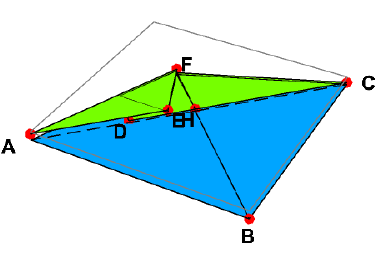}}
\quad
\subfloat[$\OO_{13}$]{\includegraphics[scale=1.2]{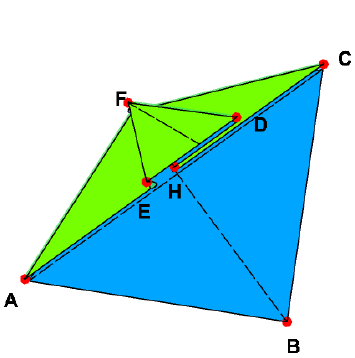}}
\qquad
\subfloat[$\OO_{14}$]{\includegraphics[scale=0.8]{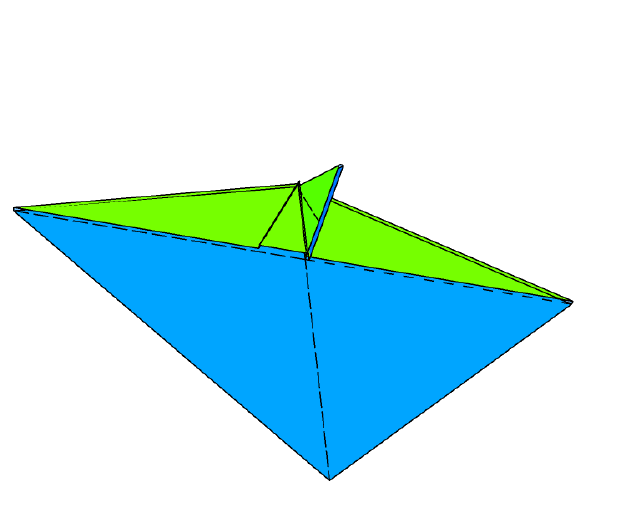}}
\caption{Rabbit ear fold}
\label{fig:rabbit-ear-fold}
\end{figure}

We show in Figs.~\ref{fig:rabbit-ear-fold}(a) - (f)  selected (geometric representation of) AOs in the construction sequence of a rabbit ear fold.  We can see an erected tab on the plane in Fig.~\ref{fig:rabbit-ear-fold}(f).  We apply Function RabbitEarFold to $\OO_{7}$ with the following arguments (inside \verb|RabbitEarFold[...]|).
\begin{verbatim}
RabbitEarFold[{11,9},{Ray["F","A"],Ray["F","H"], 
  Ray["F","D"]}]
\end{verbatim}
To be precise, the function RabbitEarFold takes a pair of faces and three rays. It makes a tab that looks like a rabbit ear standing on the isosceles triangular plane (cf.~Fig.~\ref{fig:rabbit-ear-fold}(e)). The three rays start from the same point $P$, and we represent the three rays as Ray[$P$, $Q$], Ray[$P$, $R$], and Ray[$P$, $S$].  $\triangle PQR$ is a right-angle triangle whose hypotenuse is $PQ$, and $\triangle PRS$ is another right-angle triangle whose hypotenuse is $PS$.  The function constructs rabbit ear $\triangle PRS$.
\begin{verbatim}
RabbitEarFold[{below_,above_},
 {ridge_,base_,hypotenuse_}] :=
  (CutEdge[{below,above}];
   ValleyFold[below,Rev[ridge]]; 
   ValleyFold[base];
   ValleyFold[Rev[hypotenuse]];
   ValleyFold[ridge];
   GlueEdges[$cutEdge])
    \end{verbatim}
In the above, we have Rev[Ray[$X$,$Y$]]$\defeq$Ray[$Y$, $X$] for any points $X$ and $Y$.

During the rabbit ear fold, we cut the edge FE of $\OO_{7}$.  Then, we perform four valley folds with supplied arguments.  To make the intended faces move, we need to define carefully the ray along which we perform the fold.  
The  fold of $\OO_{13}$  to  $\OO_{14}$ is a valley fold along Ray[H, F] by $\pi/2$.   The value $\pi/2$ is an additional argument of VallyFold, and it specifies the angle of rotation during the valley fold. It makes the "ear" part stand upright as shown in Fig.~\ref{fig:rabbit-ear-fold}(e).
\subsection{Pleat fold and pleat crimp fold}
\begin{figure}[h]
\center
\subfloat[]{\includegraphics[scale=0.8]{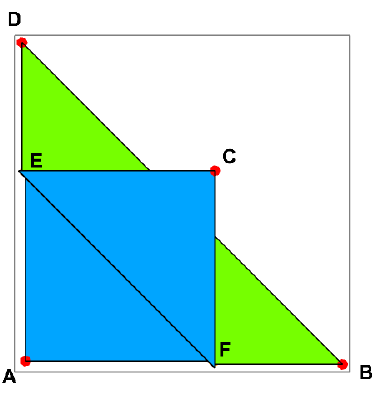}}
\quad
\subfloat[]{\includegraphics[scale=0.8]{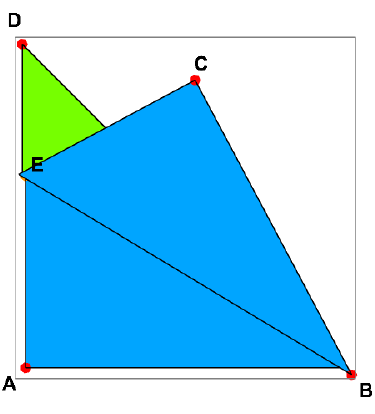}}
\quad
\subfloat[]{\includegraphics[scale=0.9]{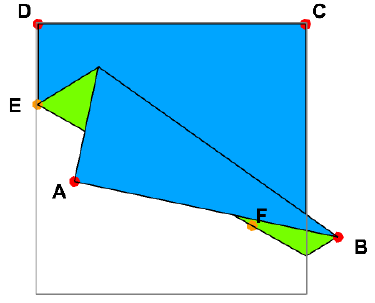}}
\caption{Examples of pleat  folds}
\label{fig:pleat-basic}
\end{figure}

A pleat is a shape pattern made of cloth or flexible flat material folded along two fold lines. In origami, a fold that makes a pleat, such as shown in Fig.~\ref{fig:pleat-basic}, is called a \emph{pleat fold}. In the figure, we only show three examples of pleat folds. The line segments DB and EF are not necessarily in parallel. We often use two lines sharing a point.  In the pleat crimp fold, we shortly explain,  we use the configuration that two line segments meet at one end. 

\begin{figure}[h]
\center
\subfloat[$\OO_{4}$]{\includegraphics[scale=1.0]{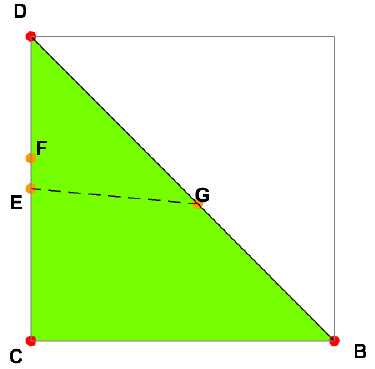}}
\quad
\subfloat[$\OO_{7}$]{\includegraphics[scale=1.0]{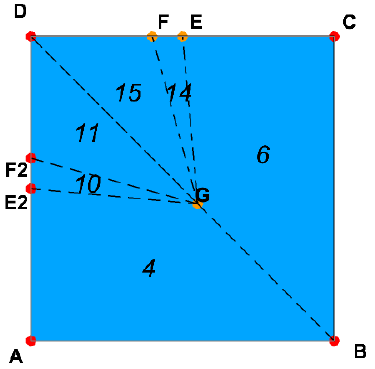}}
\quad
\subfloat[CutEdge DG: $\OO_{8}$]{\includegraphics[scale=1.0]{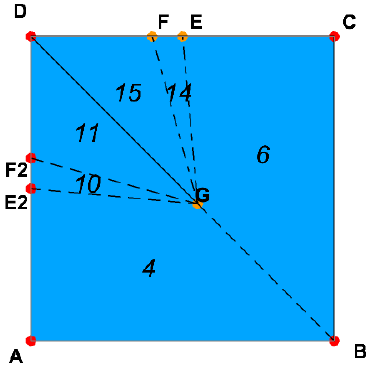}}
\quad
\subfloat[$\OO_{9}$]{\includegraphics[scale=1.0]{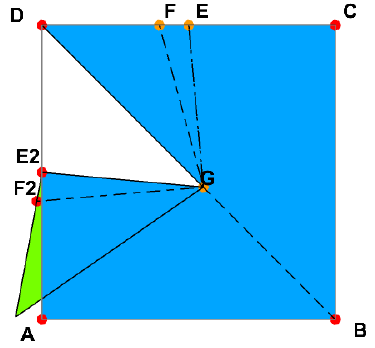}}
\quad
\subfloat[$\OO_{10}$]{\includegraphics[scale=1.0]{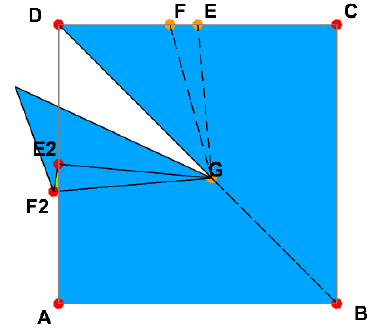}}
\quad
\subfloat[$\OO_{11}$]{\includegraphics[scale=1.0]{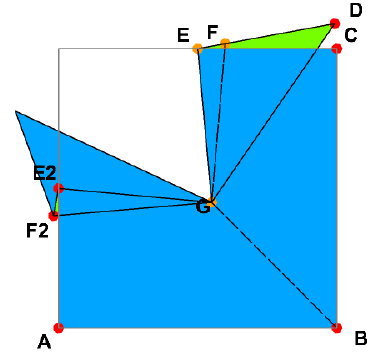}}
\quad
\subfloat[$\OO_{12}$]{\includegraphics[scale=1.0]{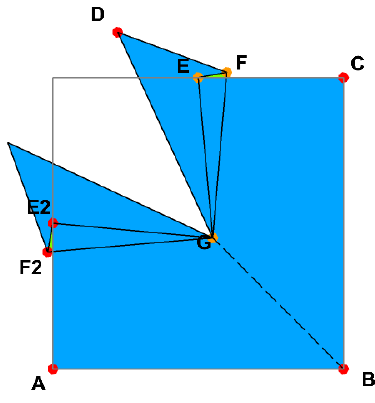}}
\quad
\subfloat[$\OO_{13}$]{\includegraphics[scale=1.0]{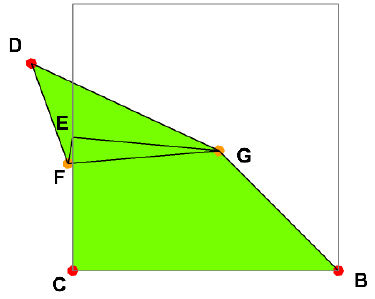}}
\quad
\subfloat[GlueEdges:$\OO_{14}$]{\includegraphics[scale=1.0]{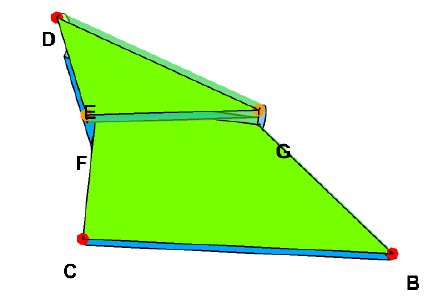}}
\caption{Pleat crimp fold with outside pleats}
\label{fig:pleat-fold}
\end{figure}

\begin{figure}[h!]
\center
\includegraphics[scale=1.6]{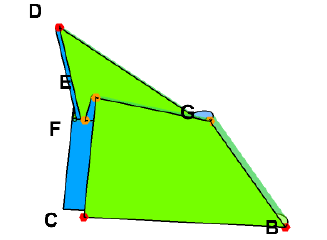}
\caption{Pleat crimp fold with inside pleats}
\label{fig:pleatcrimpinsidefold}
\end{figure}

A pleat fold is easy to perform. A pleat fold becomes more complex and exciting when combined with folds similar to an outside (or inside) reverse fold.
In Fig.~\ref{fig:pleat-fold}, we show an origami construction that makes up a pleat crimp.
The idea of realizing a pleat crimp is to divide the origami into two and apply a pleat fold to each piece. After the pleat folds on each piece, we glue them together. We have two ways of making a crimp.  The one shown in Fig.~\ref{fig:pleat-fold} has a pleat outside. 
Figure~\ref{fig:pleatcrimpinsidefold} shows the other case 
that the pleats are inside.

%
\section{Concluding remarks}
We have shown that a cut along a crease enables modeling the classical fold "squash fold" more abstract and simplifies the implementation in computational origami. Using a cut operation, we can reduce the squash fold to a sequence of valley folds and a Huzita-Justin rule.  We explained the model as an abstract origami rewrite system. Graph representation of abstract origami reveals properties among origami faces which, otherwise, would be invisible by usual origami by hand. We have also shown that we can model other popular traditional folds by cut and glue crease edges. 
 

\bibliography{adg2021}
\bibliographystyle{eptcs}

\end{document}